\documentclass[twocolumn,secnumarabic,showpacs,preprintnumbers,amsmath,amssymb,nobibnotes, aps, prl]{revtex4}

\usepackage{graphicx}
\usepackage{dcolumn}
\usepackage{bm}

\begin{document} 

\preprint{}
\input{epsf.tex}

\title{Coupling of Electronic and Motional Dynamics in a Cold Atom Optical Lattice}
\author{Hashem Zoubi and Helmut Ritsch}

\affiliation{Institut f\"{u}r Theoretische Physik, Universit\"{a}t Innsbruck, Technikerstrasse 25, A-6020 Innsbruck, Austria}  

\date{12 February, 2008}

\begin{abstract}
We study the coupling of internal electronic excitations to vibrational modes of the external motion of ultracold atoms in an optical lattice. For different ground and excited state potentials the on-site coupling of excitations and vibrations term renormalizes the effective electronic transition energy, which appears e.g. in clock transitions. In addition in the Mott state with filling factor one, the dipole-dipole coupling between neighbouring sites includes emission and absorption of vibrational quanta. Such processes create a significant mechanism for excitation of vibrations leading to motional heating of the lattice atoms by resonant light interaction. We calculate estimates of the corresponding parameters from a perturbation expansion in small atomic displacements.  
\end{abstract}


\maketitle

Since the theoretical prediction of the superfluid to the Mott insulator quantum phase transition in a system of ultracold boson atoms in an optical lattice \cite{Jaksch}, and after their experimental realization  \cite{Bloch}, the interest in such a system prolongs into new directions. The optical lattice is formed by standing waves of counter propagating laser beams with lattice constant of half laser wave length, into which the ultracold atoms are loaded. Varying the laser intensity and thus increasing the lattice depth the superfluid-Mott insulator phase transition can be achieved and is well described by the Bose-Hubbard model \cite{Jaksch}. In the Mott insulator phase a fixed number of atoms per site is obtained. For deep optical potentials the on-site potential is harmonic to a good approximation and the lowest motional states are quantized oscillator states. It is known that transitions of the atoms between these vibrational states are induced e.g. by inelastic Raman scattering of light or via scattering interactions during atom hopping among different sites \cite{Zoller}. Here we study an alternative possibility of coupling of internal atomic energy to the external motion via dipole-dipole interactions. 

For the sake of simplicity we restrict ourselves to two-level atoms close to $T=0$ in a deep optical lattice prepared in the Mott insulator phase with one atom per site. Ground and excited state optical lattice potentials are assumed to posses minima at the same positions but with different depth. In this limit vibrational excitations are localized near these minima. During the relevant time scales atoms will not hop from site to site and only interact via dipole-dipole interactions.

In previous work we already studied collective electronic excitations (excitons \cite{Zoubi}) in such a system \cite{HashemA} or its analogue with two atoms per site \cite{HashemB}. For identical ground and upper state potentials the atoms will stay localized in the lowest vibrational state, i.e. the first Bloch band. As no overlaps exist between atomic wavefunctions at different sites, an internal electronic excitation can only transfer through the lattice due to electrostatic interactions, e.g. resonance dipole-dipole interactions, where an excited atom jumps to the ground state and another atom-site is simultaneously excited. Using the lattice symmetry an excitation can be represented in the quasi-momentum space by a wave which propagates in the lattice. These energy waves are called excitons which instead of a discrete atomic energy state cover a finite exciton energy band. Scattering of such excitons off lattice defects is studied in \cite{HashemC}.

In an ideal lattice with identical upper and lower state potentials one can restrict the exciton dynamics to the lowest band. However, in more realistic optical lattices, dynamical excitation of atoms to higher vibrational states (bands) will occur. In the present work we thus include higher vibrational states in the dynamics of the collective electronic excitations. These are coupled by imperfections as e.g. differences in the upper and lower state optical potential, which results in a finite overlap between the on site wavefunctions of states with different band index. We derive estimates for this coupling between the atom internal electronic excitations and the atom external vibrational states in perturbation theory.

Due to the similarity between optical lattices in the Mott insulator phase and molecular crystals, the general formalism adopted here is in the spirit of the coupling of Frenkel excitons and optical phonons in molecular crystals \cite{Davydov,Agranovich}. In details, however, some caution is needed to identify the correct physical analogies between optical phonons in molecular crystals and the vibrational states in optical lattices. The latter are largely analogous to on-site localized acoustic phonons. In our case the interaction comprises two parts, the on-site part which gives strong excitation-vibration coupling and the transfer part, which is mediated by dipole-dipole coupling and can be treated in the perturbation theory. Still we found that this weak coupling creates an important mechanism for the emission and absorption of vibrational states.

Lets now come to a mathematical model. Effectively, the electronic excitations for ultracold atoms of an optical lattice in the Mott insulator phase can be represented by the Hamiltonian
\begin{equation}
H=\sum_i\hbar\left(\omega^e_i-\omega^g_i\right)\ B_i^{\dagger}B_i+\sum_{i,j}\hbar J_{ij}\ B_i^{\dagger}B_j,
\end{equation}
where $B_i^{\dagger},\ B_i$ are the creation and annihilation operators of an excitation at position $i$, and the $i$ summation is over the lattice sites. The internal atomic transition frequency at site $i$ is $\omega^e_i-\omega^g_i$, where we will use the approximations $\omega^{\lambda}_i=\omega^{\lambda}+D^{\lambda}({\bf R}^{\lambda}_i)$ for the ground and excited state energies $(\lambda=e,g)$. $ D^{\lambda}({\bf R}^{\lambda}_i)$ is the deviation in the internal state energy due to the on-site atomic vibration, and $\omega^{\lambda}$ is the atom vibration-free frequency. Here ${\bf R}^{\lambda}_i$ is the average shift of the atomic position at site $i$ in the $\lambda$ internal state. For small shifts it can be approximated as ${\bf R}^{\lambda}_i={\bf n}_i+{\bf u}^{\lambda}_i$, where ${\bf n}_i$ is the atomic equilibrium position in the lowest vibrational state and ${\bf u}^{\lambda}_i$ is the average deviation due to the excitation of the atom to higher vibrational states.

As shown before an internal electronic excitation can be exchanged among atoms at different sites $i$ and $j$, which is induced by dipole-dipole interactions and can be parametrized by a coupling integral $J_{ij}$. It is now essential to note that this coupling depends on the atomic shifts ${\bf u}^{\lambda}_i$ and thus on the local vibrational excitation. As long as ${\bf u}^{\lambda}_i$ is a small deviation relative to ${\bf n}_i$, which is the case for the lowest vibrational modes, to first order in this small perturbation we can split the Hamiltonian in the form $H=H_{ex}+H_{vib}+H_{ex-vib}$. Here $H_{ex}$ is the internal excitation Hamiltonian, which is obtained for atoms in the ground vibrational states, $H_{vib}$ is the atom vibration Hamiltonian, for excited and ground state atoms, and the coupling Hamiltonian $H_{ex-vib}$ is derived perturbatively for atoms excited to higher vibrational states.

To the lowest order, for atoms in the ground vibrational states, we thus get
\begin{equation}
H_{ex}=\sum_i\hbar\omega_a\ B_i^{\dagger}B_i+\sum_{i,j}\hbar J^0_{ij}\ B_i^{\dagger}B_j,
\end{equation}
where the atomic transition frequency $\omega_a=\omega^e_i-\omega^g_i$ is the same at each site, and $J^0_{ij}$ is the transfer parameter among atoms in the lowest vibrational states. The Hamiltonian can be diagonalize by using the transformation into the quasi-momentum space $B_i=\frac{1}{\sqrt{N}}\sum_{\bf k}e^{i{\bf k}\cdot{\bf n}_i}B_{\bf k}$, where $N$ is the number of sites, and ${\bf k}$ is the wave vector. We obtain the exciton Hamiltonian \cite{HashemA}
\begin{equation}
H_{ex}=\sum_{\bf k}\hbar\omega({\bf k})\ B_{\bf k}^{\dagger}B_{\bf k},
\end{equation}
where the exciton dispersion is $\omega({\bf k})=\omega_a+\sum_{\bf L}J({\bf L})e^{i{\bf k}\cdot{\bf L}}$, with ${\bf L}={\bf n}_i-{\bf n}_j$. For the case of $1D$ we have $k=(2\pi n)/(Na)$, where $a$ is the lattice constant, and $n=0,\pm 1,\cdots,\pm N/2$. For energy transfer only among nearest neighbors with parameter $J\equiv J^0$, the dispersion is given by $\omega(k)=\omega_a+2J\cos ka$. The dipole-dipole interaction between atoms at nearest neighbor sites is
\begin{equation}
\hbar J=\frac{\mu^2\left(1-3\cos^2\theta\right)}{4\pi\epsilon_0a^3},
\end{equation}
where $\mu$ is the atomic transition dipole, and $\theta$ is the angle between the transition dipole and the lattice direction.

The vibrational states of atoms at the ground and excited states are described by the harmonic  Hamiltonian
\begin{equation}
H_{vib}=\sum_i\hbar\omega_v^g\ b_i^{\dagger}b_i+\sum_i\hbar\omega_v^e\ c_i^{\dagger}c_i,
\end{equation}
where $\omega_v^g$ and $\omega_v^e$ are the vibration frequency for ground and excited state atoms, respectively. $b_i^{\dagger},\ b_i$ and $c_i^{\dagger},\ c_i$ are the creation and annihilation operators of a vibration mode at site $i$ for ground and excited state atoms, respectively. The atomic displacement operators are
\begin{equation}
\hat{u}_i^g=\sqrt{\frac{\hbar}{2m\omega_v^g}}\left(b_i+b_i^{\dagger}\right)\ ,\ \hat{u}_i^e=\sqrt{\frac{\hbar}{2m\omega_v^e}}\left(c_i+c_i^{\dagger}\right),
\end{equation}
where $m$ is the atomic mass.

To first order in the perturbation series with respect to ${\bf u}^{\lambda}_i$, we get the excitation-vibration coupling by the Hamiltonian $H_{ex-vib}=H_{ex-vib}^I+H_{ex-vib}^{II}$, where $H_{ex-vib}^I$ is for the on-site part, and $H_{ex-vib}^{II}$ for the transfer part. The on-site part is given by
\begin{equation}
H_{ex-vib}^I=\sum_i\hbar\left[M^e_i\left(c_i+c_i^{\dagger}\right)-M^g_i\left(b_i+b_i^{\dagger}\right)\right]\ B_i^{\dagger}B_i,
\end{equation}
where the coupling parameter is
\begin{equation}
M^{\lambda}_i=\sqrt{\frac{\hbar}{2m\omega_v^{\lambda}}}\left\{\frac{\partial D^{\lambda}({\bf R}^{\lambda}_i)}{\partial {\bf u}^{\lambda}_i}\right\}_{{\bf u}^{\lambda}_i=0},
\end{equation}
which is related to the slope of $D^{\lambda}({\bf R}^{\lambda}_i)$. As $D^{\lambda}({\bf R}^{\lambda}_i)$ is a function of the atomic wave function at site $i$ and in a fixed Bloch band, the parameter $M^{\lambda}_i$ is related to the correlations between atomic wave functions at the initial and the final Bloch bands. The transfer part is given by
\begin{eqnarray}
H_{ex-vib}^{II}&=&\sum_{i,j}\hbar\left[F^{ei}_{ij}\ c_i^{\dagger}+F^{gi}_{ij}\ b_i\right. \nonumber \\
&+&\left.F^{ej}_{ij}\ c_j+F^{gj}_{ij}\ b_j^{\dagger}\right]\ B_i^{\dagger}B_j,
\end{eqnarray}
where the coupling parameter is
\begin{equation} \label{TranCoup}
F^{\lambda i}_{ij}=\sqrt{\frac{\hbar}{2m\omega_v^{\lambda}}}\left\{\frac{\partial J_{ij}}{\partial {\bf u}^{\lambda}_i}\right\}_{{\bf u}^{\lambda}_i=0},
\end{equation}
which is related to the deviation in $J_{ij}$.

In the strong on-site excitation-vibration coupling regime, namely in the limit of $M_i^{\lambda}\gg F^{\lambda i}_{ij}$, we can apply the canonical transformation \cite{Mahan} $\tilde{\cal O}=e^{\hat{\sigma}}{\cal O}e^{-\hat{\sigma}}$, where $\hat{\sigma}=\hat{s}\ B_i^{\dagger}B_i$, with $\hat{s}=\left[\frac{M_i^g}{\omega_v^g}\left(b_i^{\dagger}-b_i\right)-\frac{M_i^e}{\omega_v^e}\left(c_i^{\dagger}-c_i\right)\right]$, to get $\tilde{b}_i=b_i-\frac{M_i^g}{\omega_v^g}\ B_i^{\dagger}B_i$ and $\tilde{c}_i=c_i+\frac{M_i^e}{\omega_v^e}\ B_i^{\dagger}B_i$, and with $\tilde{B}_i=B_i\hat{X}$, where $\hat{X}=e^{-\hat{s}}$. The new excitation represents an electronic excitation dressed by a cloud of on-site vibrations and can be considered as an excitation-polaron. Furthermore, we assume the excitation-vibration coupling parameters to be site independent, by defining $M_{\lambda}\equiv M_i^{\lambda}$ and $F^{\lambda}_{ij}\equiv F^{\lambda i}_{ij}$. In terms of the new operators, the excitation Hamiltonian reads
\begin{equation}
H_{ex}=\sum_i\hbar\omega_0\ \tilde{B}_i^{\dagger}\tilde{B}_i+\sum_{i,j}\hbar J^0_{ij}\ \tilde{B}_i^{\dagger}\tilde{B}_j,
\end{equation}
where $\omega_0=\omega_a-\Delta$, with $\Delta=\frac{M^{g\ 2}}{\omega_v^g}+\frac{M^{e\ 2}}{\omega_v^e}$. In this limit the effect of the on-site excitation-vibration coupling is to renormalize the excitation transition energy by an energy shift of $\Delta$, where $\omega_a\gg \Delta$, and which needs to be considered in the clock state transition energy \cite{Katori}. The vibration Hamiltonian is now given by $H_{vib}=\sum_i\hbar\omega_v^g\ \tilde{b}_i^{\dagger}\tilde{b}_i+\sum_i\hbar\omega_v^e\ \tilde{c}_i^{\dagger}\tilde{c}_i$. The excitation-vibration coupling Hamiltonian of the transfer part, which is considered as a perturbation part, is written as
\begin{equation}
H_{ex-vib}=\sum_{i,j}\hbar\left[F^{e}_{ij}\left(\tilde{c}_i^{\dagger}+\tilde{c}_j\right)+F^{g}_{ij}\left(\tilde{b}_i+\tilde{b}_j^{\dagger}\right)\right]\ \tilde{B}_i^{\dagger}\tilde{B}_j.
\end{equation}
This term describes emission and absorption of vibrations, and which induced by excitation transfer among different sites. The four processes are plotted in figures (1-4). The processes show emission and absorption of a vibration due to the excitation transfer among nearest neighbor sites. Process I show emission of ground state vibration at site $j$, and process II emission of excited state vibration at site $i$. While process III show absorption of ground state vibration at site $i$, and process IV absorption of excited state vibration at site $j$. The discussion is limited here to the first order of the perturbation series, where the emission and absorption processes are between the ground and the first excited vibrational state with a single vibrational quanta. Higher orders include processes with more than a single vibrational quanta.
\begin{figure}
\centerline{\epsfxsize=6.0cm \epsfbox{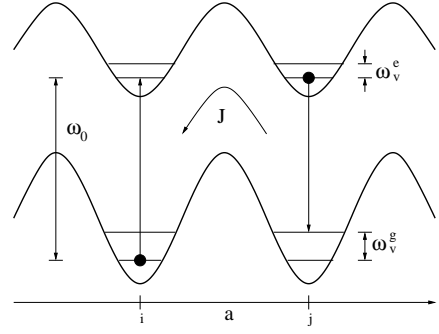}}
\caption{Process I for $F^{g}_{ij}\ \tilde{B}_i^{\dagger}\tilde{B}_j\ \tilde{b}_j^{\dagger}$, emission of ground state vibration at site $j$.}
\end{figure}
\begin{figure}
\centerline{\epsfxsize=6.0cm \epsfbox{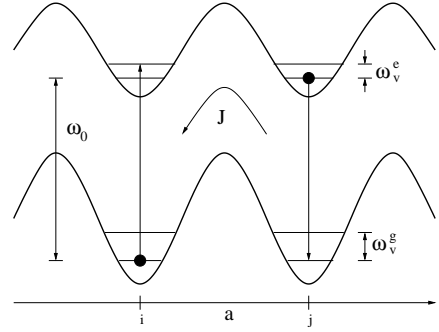}}
\caption{Process II for $F^{e}_{ij}\ \tilde{B}_i^{\dagger}\tilde{B}_j\ \tilde{c}_i^{\dagger}$, emission of excited state vibration at site $i$.}
\end{figure}
\begin{figure}
\centerline{\epsfxsize=6.0cm \epsfbox{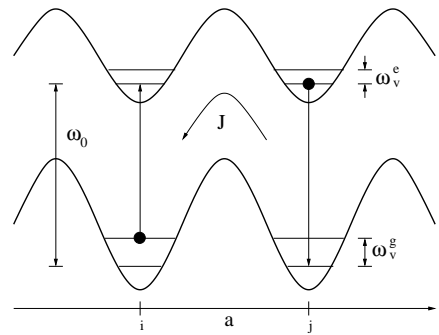}}
\caption{Process III for $F^{g}_{ij}\ \tilde{B}_i^{\dagger}\tilde{B}_j\ \tilde{b}_i$, absorption of ground state vibration at site $i$.}
\end{figure}
\begin{figure}
\centerline{\epsfxsize=6.0cm \epsfbox{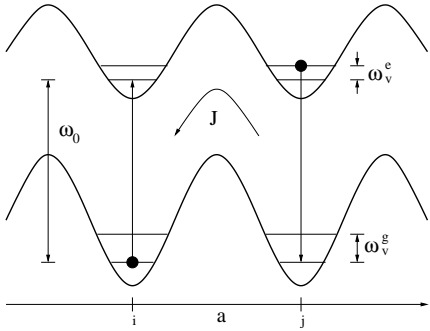}}
\caption{Process IV for $F^{e}_{ij}\ \tilde{B}_i^{\dagger}\tilde{B}_j\ \tilde{c}_j$, absorption of excited state vibration at site $j$.}
\end{figure}

The excitation-vibration coupling energy for the case of dipole-dipole interaction in $1D$, and for only nearest neighbors, is given by
\begin{equation}
\hbar F^{\lambda}=\sqrt{\frac{\hbar}{2m\omega_v^{\lambda}}}\ \frac{3\mu^2\left(3\cos^2\theta-1\right)}{4\pi\epsilon_0a^4}.
\end{equation}
In figure (5) we plot the energy transfer $\hbar J$ and the above coupling $\hbar F$ as a function of the angle $\theta$. We used the typical numbers $\mu=2\ e\AA$, $a=2000\ \AA$, $mc^2=10^{12}\ eV$, and $\hbar\omega_v=10^{-9}\ eV$. In this case we have $J\gg F$, and they have an opposite sign, with a crossover at $\theta\approx 54.7$. The ratio between the two parameters is $F/J=3\bar{a}/a$, with the vibration average length $\bar{a}=\sqrt{\frac{\hbar}{2m\omega_v}}$, where here we have $F/J\approx 0.1$. For the emission and absorption processes through the excitation transfer to take place, the excitation transfer time needs to be larger than the excited state life time.
\begin{figure}
\centerline{\epsfxsize=6.0cm \epsfbox{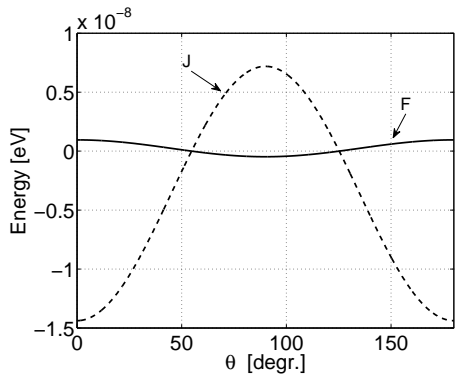}}
\caption{The energy transfer $\hbar J$ and the coupling $\hbar F$ vs. the angle $\theta$.}
\end{figure}

The Hamiltonian in terms of transformed operators can be cast into the momentum space representation. The excitation Hamiltonian, as before, casts into the exciton Hamiltonian $H_{ex}=\sum_{\bf k}\hbar\omega({\bf k})\ \tilde{B}_{\bf k}^{\dagger}\tilde{B}_{\bf k}$. The vibration Hamiltonian, even though represents on-site localized vibrations, can be transformed formally into the momentum space, by applying the transformations $\tilde{b}_i=\frac{1}{\sqrt{N}}\sum_{\bf q}e^{i{\bf q}\cdot{\bf n}_i}\tilde{b}_{\bf q}$, and $\tilde{c}_i=\frac{1}{\sqrt{N}}\sum_{\bf q}e^{i{\bf q}\cdot{\bf n}_i}\tilde{c}_{\bf q}$, to get $H_{vib}=\sum_{\bf q}\hbar\omega_v^g\ \tilde{b}_{\bf q}^{\dagger}\tilde{b}_{\bf q}+\sum_{\bf q}\hbar\omega_v^e\ \tilde{c}_{\bf q}^{\dagger}\tilde{c}_{\bf q}$, which have flat dispersions, namely $q$-independent. The exciton-vibration coupling now reads
\begin{eqnarray}
H_{ex-vib}&=&\sum_{\bf k,q}\hbar\left\{F^{e}({\bf k+q})\ \tilde{c}_{\bf q}+F^{g}({\bf k})\ \tilde{b}_{\bf q}\right. \nonumber \\
&+&\left.F^{e}({\bf k})\ \tilde{c}_{\bf -q}^{\dagger}+F^{g}({\bf k+q})\ \tilde{b}_{\bf -q}^{\dagger}\right\}\ \tilde{B}_{\bf k+q}^{\dagger}\tilde{B}_{\bf k},\nonumber \\
\end{eqnarray}
where $F^{\lambda}({\bf k})=\frac{1}{\sqrt{N}}\sum_{\bf L}F^{\lambda}({\bf L})e^{i{\bf k}\cdot{\bf L}}$. In assuming transfer among only nearest neighbor sites we get $F^{\lambda}({\bf k})=\frac{2}{\sqrt{N}}F^{\lambda}\cos ka$. The Hamiltonian $H_{ex-vib}$ describes scattering of excitons between different wave vectors by the emission and absorption of a vibration. Such a process is possible in the limit of $J\gg \omega_v^{\lambda}$, where the exciton band width is larger than the vibration energy. The scattering conserves energy and momentum, and as the vibration dispersion is a flat one, the vibration can absorb any momentum amount of the exciton. The exciton transition rate off a spontaneous emission is $w^{\lambda}_{\bf k}=\frac{2\pi}{\hbar}\ |\hbar F^{\lambda}({\bf k})|^2$, which results of the Fermi golden rule.

In the case in which the ground and excited state optical lattice potentials are approximately identical, we have almost identical ground and excited vibrational states, then we have $ D^{e}({\bf R}^{e}_i)-D^{g}({\bf R}^{g}_i)\approx 0$. Now, the on-site and the transfer parts of the excitation-vibration coupling need to be treated as perturbations, and in the limit of $M_i^{\lambda}\ll F^{\lambda i}_{ij}$ we can consider only the transfer part. In this case no significant on-site excitation-vibration coupling exists, and no shift in the transition energy is obtained. We can use exactly the previous derivation for the transfer part of the excitation-vibration coupling.

In a summary we analysed a new mechanism for the coupling of different vibrational states in optical lattices, mediated via internal electronic excitations. Using perturbation theory an effective explicit form of this excitation-vibration interaction is derived and shows that this coupling can be stronger than other known mechanisms via tunneling and scattering interactions. The on-site part of the excitation-vibration coupling is treated in the strong coupling regime, and results in a renormalization of the atomic transition energy, which is of importance for clock states. While, the energy transfer part is treated in the perturbation theory, and results in the emission and absorption of a single vibrational quanta accompanied by the excitation transfer among nearest neighbor sites. While the mechanism is most prominent for resonant light interaction as used in atomic lattice clocks, it will remain present even for stronger detuned light, which is often used to manipulate and control atomic dynamics in such lattices.  Naturally  the exciton-vibration interaction will serve as an important source for relaxation of collective electronic excitations (excitons) towards lower energies of the exciton band and thermal equilibrium in optical lattices.

The work was supported by the Austrian Science Funds (FWF), via the Lise-Meitner Program (M977).

\end{document}